\documentclass{article}
\usepackage{graphicx}

\usepackage[final]{neurips_2023}

\usepackage[utf8]{inputenc} 
\usepackage[T1]{fontenc}    
\usepackage{hyperref}       
\usepackage{url}            
\usepackage{booktabs}       
\usepackage{amsfonts}       
\usepackage{nicefrac}       
\usepackage{microtype}      
\usepackage{xcolor}         
\usepackage{graphicx}
\usepackage{subfig}
\usepackage{multirow}

\title{Cancer-Net PCa-Gen: Synthesis of Realistic Prostate Diffusion Weighted Imaging Data via Anatomic-Conditional Controlled Latent Diffusion}

\author{%
  Aditya Sridhar\\
  Department of Systems Design Engineering\\
  University of Waterloo \\
  \And
  Chi-en Amy Tai \\
  Department of Systems Design Engineering\\
  University of Waterloo\\
  \And
  Hayden Gunraj\\
  Department of Systems Design Engineering\\
  University of Waterloo \\
  \And
  Yuhao Chen\\
  Department of Systems Design Engineering\\
  University of Waterloo \\
  \And
  Alexander Wong \\
  Department of Systems Design Engineering\\
  University of Waterloo \\
}

\begin{document}

\maketitle

\begin{abstract}
In Canada, prostate cancer is the most common form of cancer in men and accounted for 20\% of new cancer cases for this demographic in 2022. Due to recent successes in leveraging machine learning for clinical decision support, there has been significant interest in the development of deep neural networks for prostate cancer diagnosis, prognosis, and treatment planning using diffusion weighted imaging (DWI) data. A major challenge hindering widespread adoption in clinical use is poor generalization of such networks due to scarcity of large-scale, diverse, balanced prostate imaging datasets for training such networks. In this study, we explore the efficacy of latent diffusion for generating realistic prostate DWI data through the introduction of an anatomic-conditional controlled latent diffusion strategy. To the best of the authors’ knowledge, this is the first study to leverage conditioning for synthesis of prostate cancer imaging. Experimental results show that the proposed strategy, which we call Cancer-Net PCa-Gen, enhances synthesis of diverse prostate images through controllable tumour locations and better anatomical and textural fidelity. These crucial features make it well-suited for augmenting real patient data, enabling neural networks to be trained on a more diverse and comprehensive data distribution. The Cancer-Net PCa-Gen framework and sample images have been made publicly available at \footnote{https://www.kaggle.com/datasets/deetsadi/cancer-net-pca-gen-dataset} as a part of a global open-source initiative dedicated to accelerating advancement in machine learning to aid clinicians in the fight against cancer.
\end{abstract}

\section{Introduction}
In Canada, prostate cancer is the most common form of cancer in men~\cite{CanadaGOV} and accounted for 20\% of new cancer cases for this demographic in 2022~\cite{CancerCA}. Furthermore, an estimated 13 Canadian men will die from prostate cancer every day, with an estimated 4,600 total deaths in 2022~\cite{CancerCA}. Neoteric methods of applying deep neural networks for assisting clinicians in the detection and treatment planning of prostate cancer using diffusion weighted imaging (DWI) have shown strong results~\cite{yoo2019prostate}, but are precluded from widespread adoption in the medical field due to concerns over their generalization. This is rooted in limited training data for these networks which poorly represents the expansive distribution of patient scans. Recent advances in generative image models, especially those applying a latent diffusion process~\cite{rombach2022highresolution}, have inspired ideas of finetuning foundation models for medical data generation~\cite{stanford}. Most notably, previous work by Khader et. al.~\cite{khader2023medical} explores training Denoising Diffusion Probabilistic Models (DDPMs)~\cite{ho2020denoising} on Knee MRIs from the MRNet dataset~\cite{bien2018deep}. However, the stochastic nature of DDPMs leads to uncontrollable results, which can cause visually corrupted data through incorrect anatomy and structure. Hence, the performance of neural networks trained using this data can be unpredictable and are therefore unsuitable for medical decision making. 

Motivated by challenges faced when generating synthetic training data that represents a real patient distribution, we introduce Cancer-Net PCa-Gen, an anatomic-conditional controlled latent diffusion strategy that generates realistic prostate DWI data. In order to demonstrate the efficacy of our network designs, we utilize the publicly available Cancer-Net PCa~\cite{Wong2022} dataset, which consists of 200 prostate scans collected from patients at the Prostate MRI Reference Center in Nijmegen, The Netherlands. To the best of the authors’ knowledge, this is the first study to leverage conditioning for synthesis of prostate cancer imaging. Experimental results of Cancer-Net PCa-Gen with inputs of a single DWI channel and with inputs of three DWI channels show that it is capable of improved synthesis of diverse prostate images with controllable tumour locations and better anatomical and textural fidelity, making it well-suited for augmenting real patient data. This enables neural networks to be trained on a more diverse and comprehensive data distribution. The Cancer-Net PCa-Gen framework and sample images have been made publicly available at \footnote{https://www.kaggle.com/datasets/deetsadi/cancer-net-pca-gen-dataset} as a part of a global open-source initiative dedicated to accelerating advancement in machine learning to aid clinicians in the fight against cancer.

\section{Methodology}
\begin{figure}[h]
    \centering
    \includegraphics[width=\linewidth]{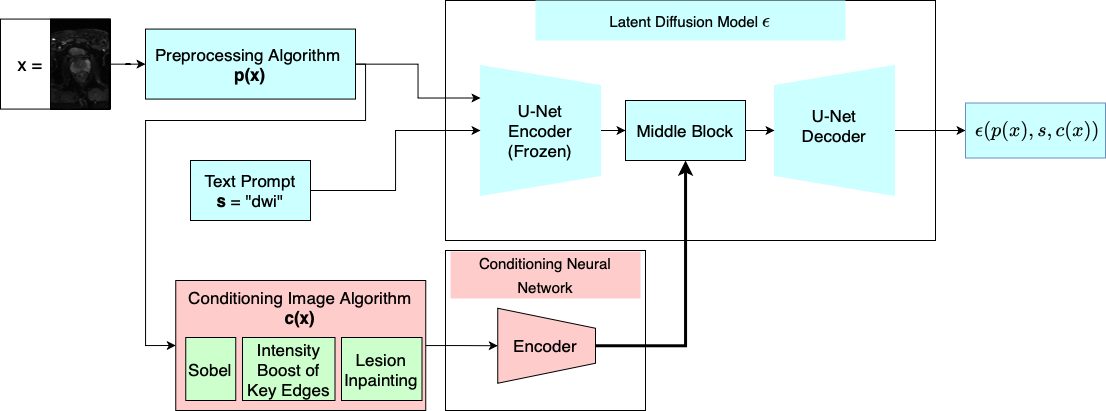}
    \caption{A high-level view demonstrating the proposed anatomic-conditional controlled latent diffusion framework.}
    \label{fig:process-view}
\end{figure}

Our work utilizes the Cancer-Net PCa-Data~\cite{Wong2022} dataset for patient DWI scans. To ensure texture fidelity in our generated images, we design a novel data processing pipeline to enhance features of the DWI data which most prominently impact output quality for our base text-to-image network. A high-level view of the framework is shown in Figure~\ref{fig:process-view}. As seen, there are two main components for Cancer-Net PCa-Gen: the latent diffusion model (blue) and the conditioning neural network (red). After training the base latent diffusion model, the encoder of the U-Net is frozen and the output of the conditioning neural network is fed into the latent diffusion model (middle block) to generate the final image.

\begin{figure}[t]
    \centering
    \includegraphics[width=\linewidth]{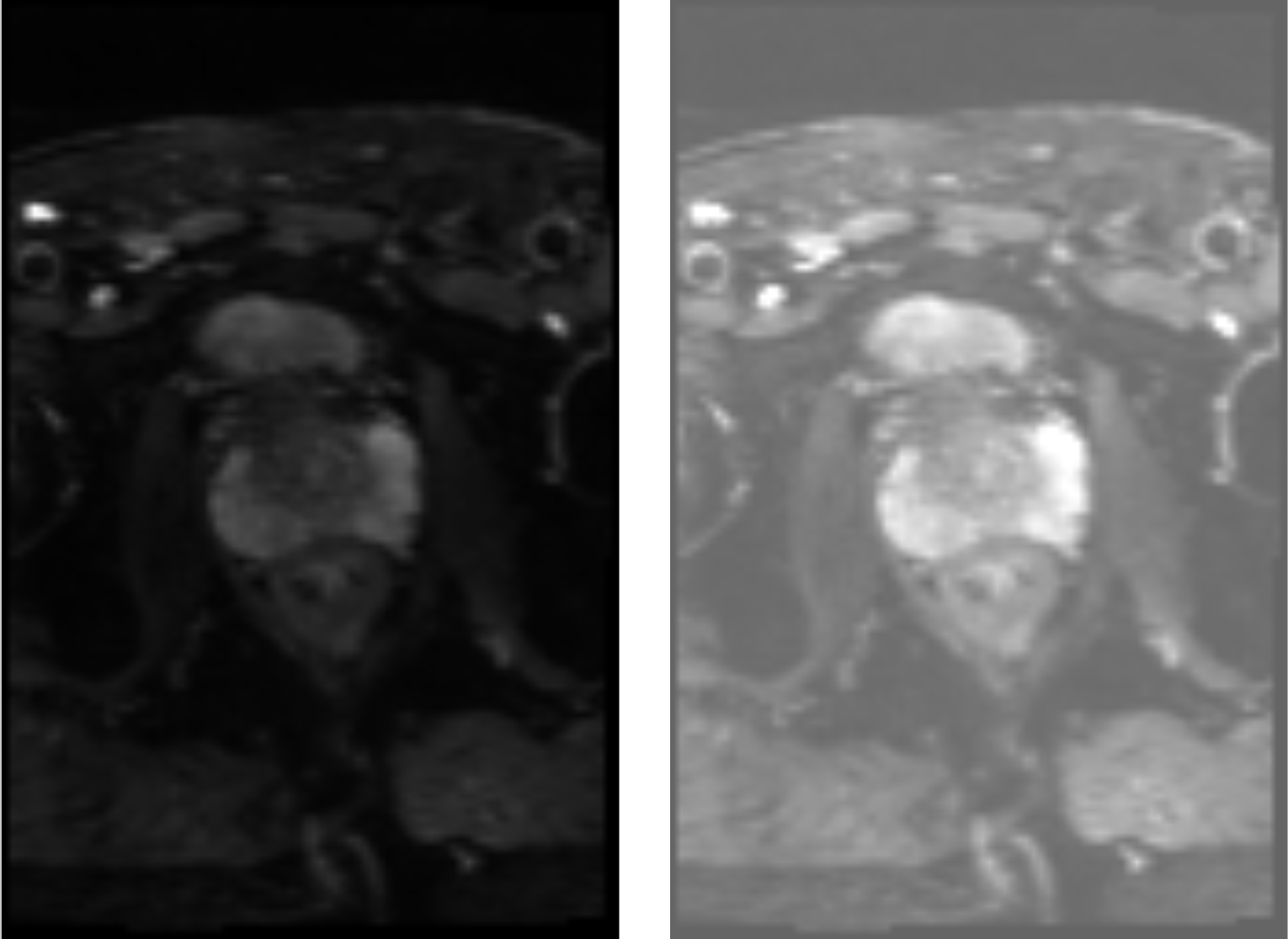}
    \caption{Example of an original middle slice DWI (b=50) sample (left) and the enhanced image produced after passing it through \(p(x)\) (right).}
    \label{fig:preprocessed_and_ground_truth_image_comparison}
\end{figure}

For each example \(x\) in our training data, where $x \in \mathbb{R}^{128 \times 84}$ for single channel DWI (b=50) and $x \in \mathbb{R}^{128 \times 84 \times 3}$ for three channel DWI (b=50, 400, 800), we pass \(x\) through our preprocessing function \(p(x)\) defined as
$$p(x) = (x - (mean(x) - 4 * std(x))) / (8 * std(x))$$, where $mean(x)$ and $std(x)$ are the mean and standard deviation functions respectively. Note that both $mean(x)$ and $std(x)$ produce scalar values, and $p(x)$ produces output of same dimensionality as $x$. 
The preprocessing function enhances the effectiveness of our latent diffusion model in identifying and reproducing the structure and texture of training samples. An example of this is shown in Figure-\ref{fig:preprocessed_and_ground_truth_image_comparison}.

\begin{figure}[t]
    \centering
    \includegraphics[width=\linewidth]{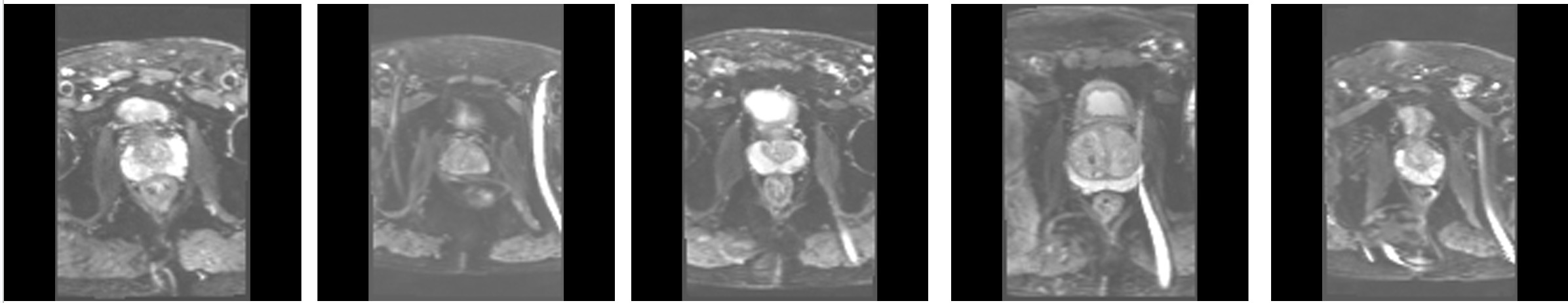}
    \caption{Example middle slices from DWI samples (b=50) after preprocessing.}
    \label{fig:ground-truth-dwi}
\end{figure}

After preprocessing using $p(x)$, the images are resized and padded to shape 512$\times$512$\times$3. Examples of these images with enhanced image clarity are shown in Figure~\ref{fig:ground-truth-dwi}. Since our anatomic-conditional controlled latent diffusion strategy enables stronger fine-grained control of image structure compared to a textual description, we specify the text prompt, \(s\), as the string constant \(``dwi"\).

\begin{figure}[t]
    \centering
    \includegraphics[width=\linewidth]{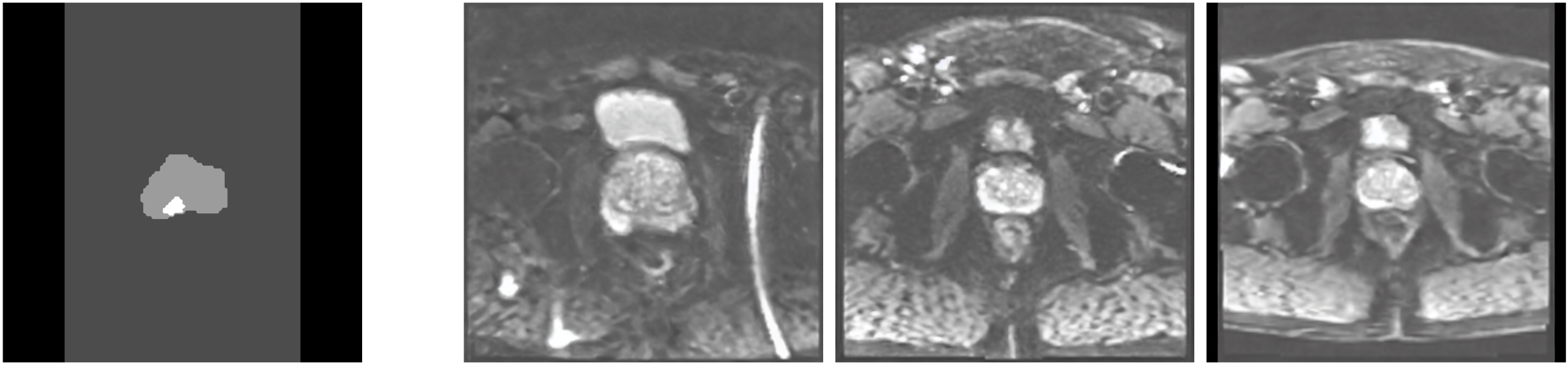}
    \caption{Images from initial approach where the generated images (right) cannot be guided by the single generated mask (left).}
    \label{fig:failed-initial-images}
\end{figure}

To construct the anatomic-conditional component of our framework and control the latent diffusion model, our initial approach of utilizing ground truth segmentation masks containing only the prostate and lesion failed to control the anatomy of output images (see Figure~\ref{fig:failed-initial-images}). Subsequently, we employ a custom conditioning algorithm $c(x)$ to generate anatomical maps of  DWI scans, which are passed to the model alongside the original scan image when training.

\begin{figure}[!ht]
    \centering
    \subfloat[Ground Truth]{\includegraphics[width=.2\linewidth]{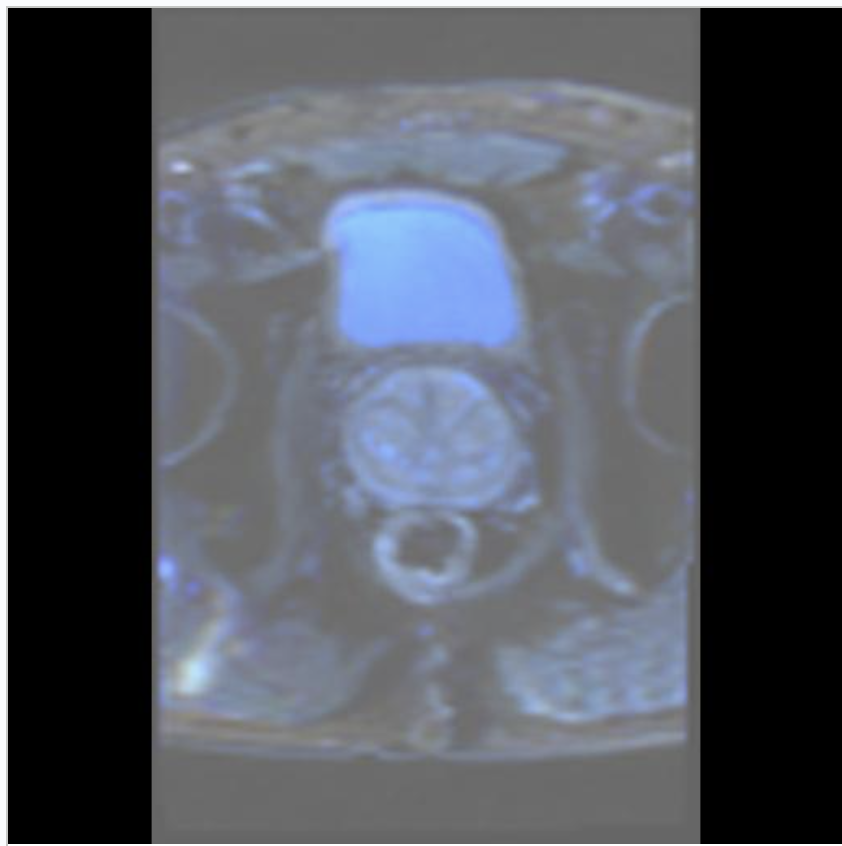}}
    \hfil
    \subfloat[Anatomical Mask]{\includegraphics[width=.2\linewidth]{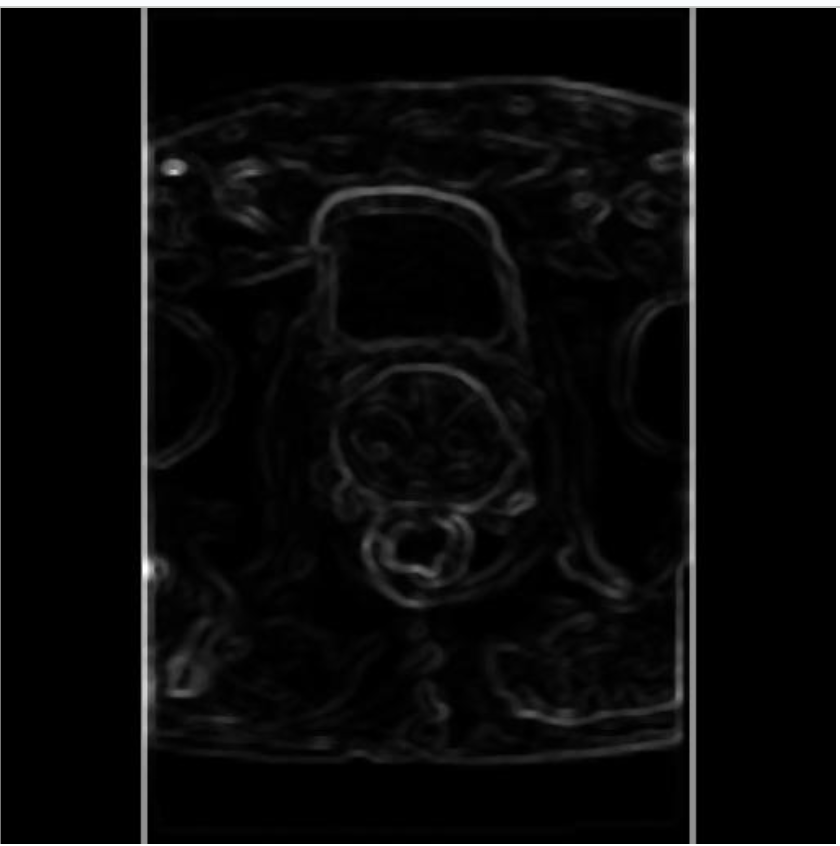}}
    \hfil
    \subfloat[Overlay]{\includegraphics[width=.2\linewidth]{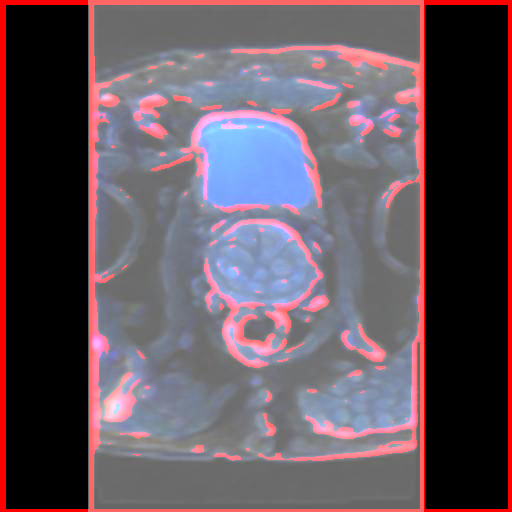}}
    \hfil
    \caption{A training sample pair of the preprocessed ground truth DWI $x_s$ with all three channels (b=50, 400, 800) stacked onto a single image (a), its associated anatomical conditioning mask $x_m$ (b), and an overlay of the anatomical conditioning mask on the ground truth image (c).}
    \label{fig:ground-truth-and-mask}
\end{figure}

Specifically, we construct $c(x)$ as a multifaceted tool to emphasize anatomic components, consisting of a Sobel operator~\cite{kanopoulos1988design} for edge detection, intensity boosting of crucial structural elements in the prostate (such as key edges), and inpainting a lesion outline. To prepare a scan image $x_s$ and anatomical mask $x_m$ pair for training, the original DWI training sample \(x\) is first preprocessed using $p(x)$ to get $x_s$, and $x_s$ is then passed to $c(x)$ to generate $x_m$, an anatomical mask with the same image dimensions as $x_s$. An example of a sample and conditioning pair is shown in Figure~\ref{fig:ground-truth-and-mask}.

To train the framework, we leverage a novel suite-based method for data synthesis by applying a latent diffusion~\cite{rombach2022highresolution} model \(\epsilon\) to learn a high-level representation of DWI data, and then using a fine-grained conditioning image strategy based on previous work by Zhang et. al.~\cite{zhang2023adding} for controlling diffusion models. Specifically, we design the base latent diffusion model to synthesize data containing the structure and texture of the prostate region using the data preprocessed through $p(x)$ and fixed text prompt $s$. Further, we build our anatomical-conditional networks through freezing the encoder of our base latent diffusion model and training an auxiliary neural network to inject conditioning information, which is then used in the decoding process. These networks utilize the pairwise training examples, consisting of a ground truth DWI scan and its corresponding conditioning image, to learn how to guide the generated scan features based on the conditioning image. We train latent diffusion models on full-sized and cropped data for both a single DWI channel (b=50) and across all three DWI channels (b=50, 400, 800) to ensure that ground truth features are learned most accurately in all cases. Since the Cancer-Net PCa Dataset contains only 200 data samples, we utilize all of them when training both the latent diffusion and auxiliary networks. 

\section{Results}
We first investigate the base latent diffusion model without the anatomic-conditional image algorithm and then the results with the anatomic-conditional image algorithm for both the single channel (b=50) and across multiple channels (b=50, 400, 800). Notably we use a simpler conditioning image algorithm $c_s(x)$ for the single channel as it is similarly effective but requires less steps. We design $c_s(x)$ to contain only prostate and lesion outlines, without the edge detection and intensity manipulation present in $c(x)$. However, $c_s(x)$ fails for the typical multiple channel DWI generation use case and instead, $c(x)$ should be used as in training.

\subsection{Latent Diffusion}
\begin{figure}[!ht]
    \centering
    \subfloat[Original]{\includegraphics[width=.205\linewidth]{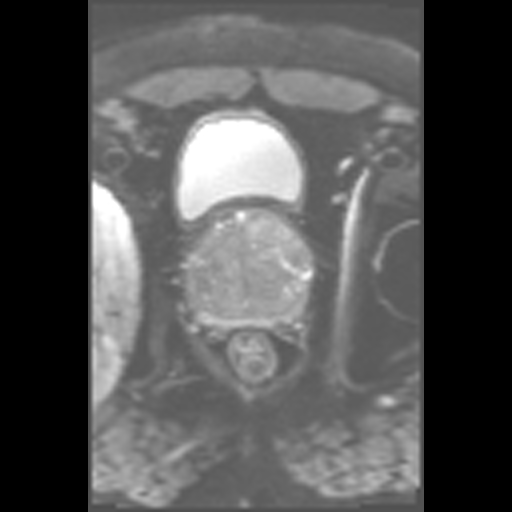}}
    \hfil
    \subfloat[Generated Images]{\includegraphics[width=.8\linewidth]{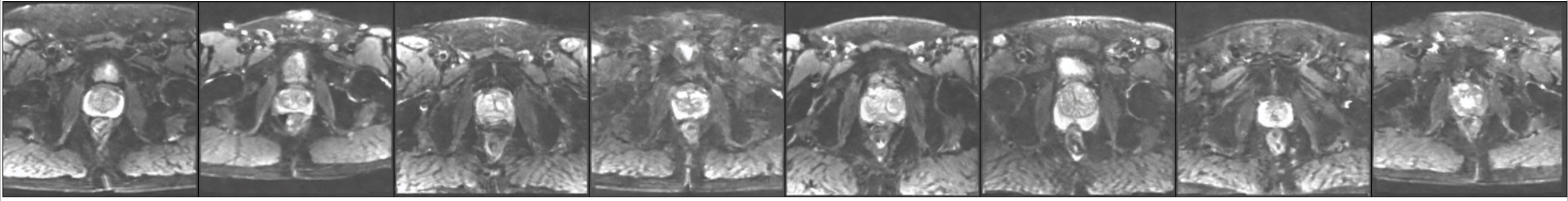}}
    \caption{Generated data samples from our latent diffusion model trained on DWI (b=50) slices.}
    \label{fig:generated-dwi}
\end{figure}

\begin{figure}[!ht]
    \centering
    \includegraphics[width=\linewidth]{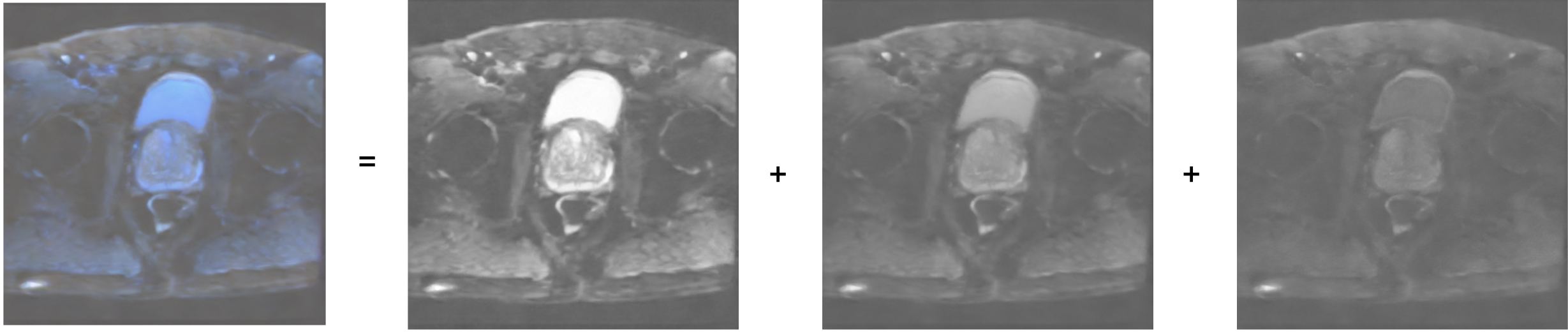}
    \caption{Channel-wise decomposition of synthesized DWI (b=50, 400, 800) image.}
    \label{fig:generated-dwi-decomp}
\end{figure}

We evaluate the efficacy of our base latent diffusion model through generating images with the same text prompt \(s\) used during training. Examples of generated data samples for the model trained on the single channel (b=50) including all relevant anatomic components and texture fidelity is shown in Figure~\ref{fig:generated-dwi}. We also demonstrate the ability to reproduce ground truth like textures and anatomic components, such as the bladder, as well as channel-wise scan characteristics for DWI (b=50, 400, 800). The channel-wise decomposition of synthesized DWI across all three channels is shown in Figure~\ref{fig:generated-dwi-decomp} and the decaying intensity over successive channels showcases the network modelling increasing b-value characteristics.

\subsection{Anatomic-Conditional Generation}
\begin{figure}[!ht]
    \centering
    \subfloat[DWI (b=50)]{\includegraphics[width=.21\linewidth]{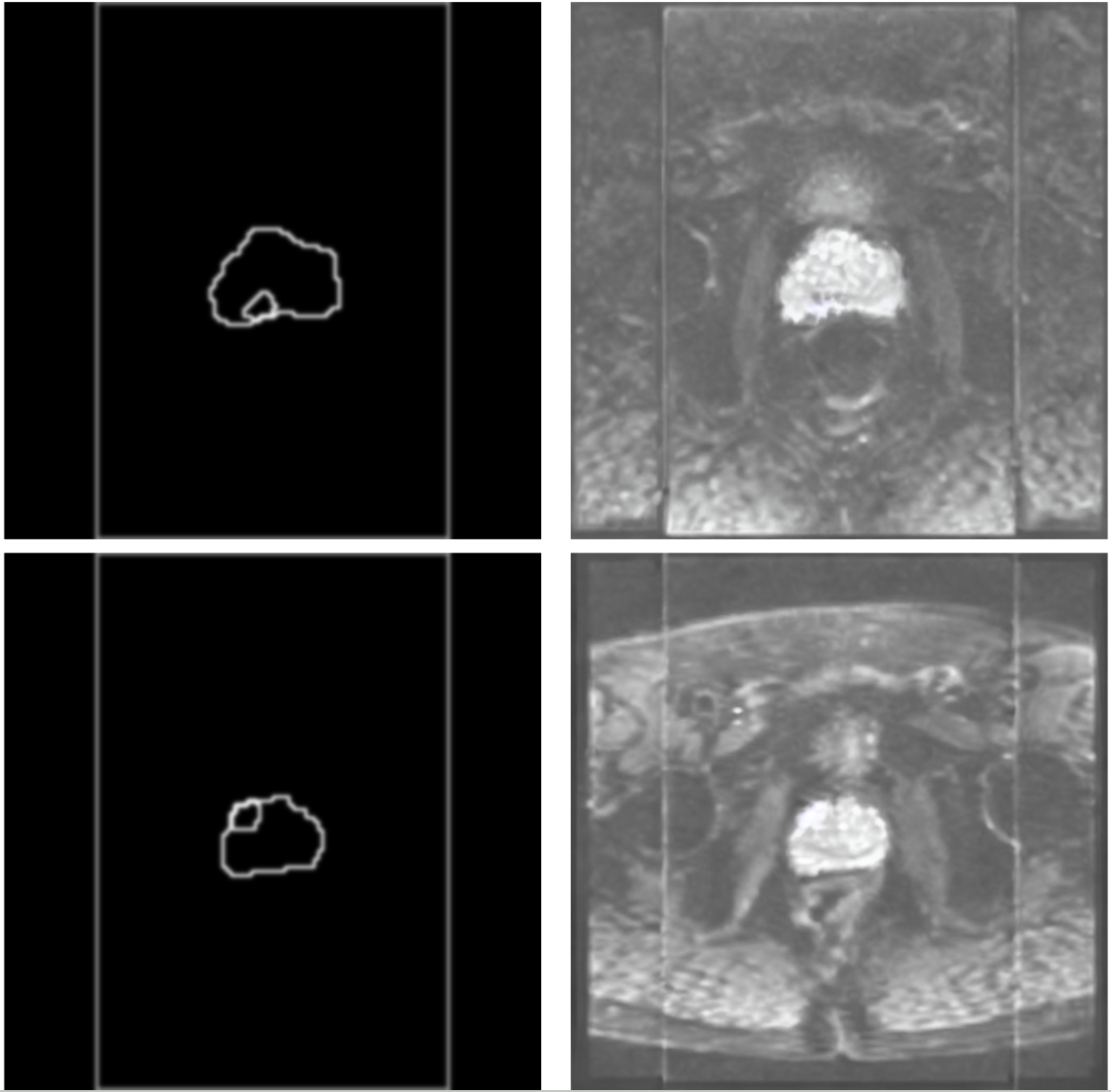}}
    \hfil
    \subfloat[DWI (b=50,400,800)]{\includegraphics[width=.22\linewidth]{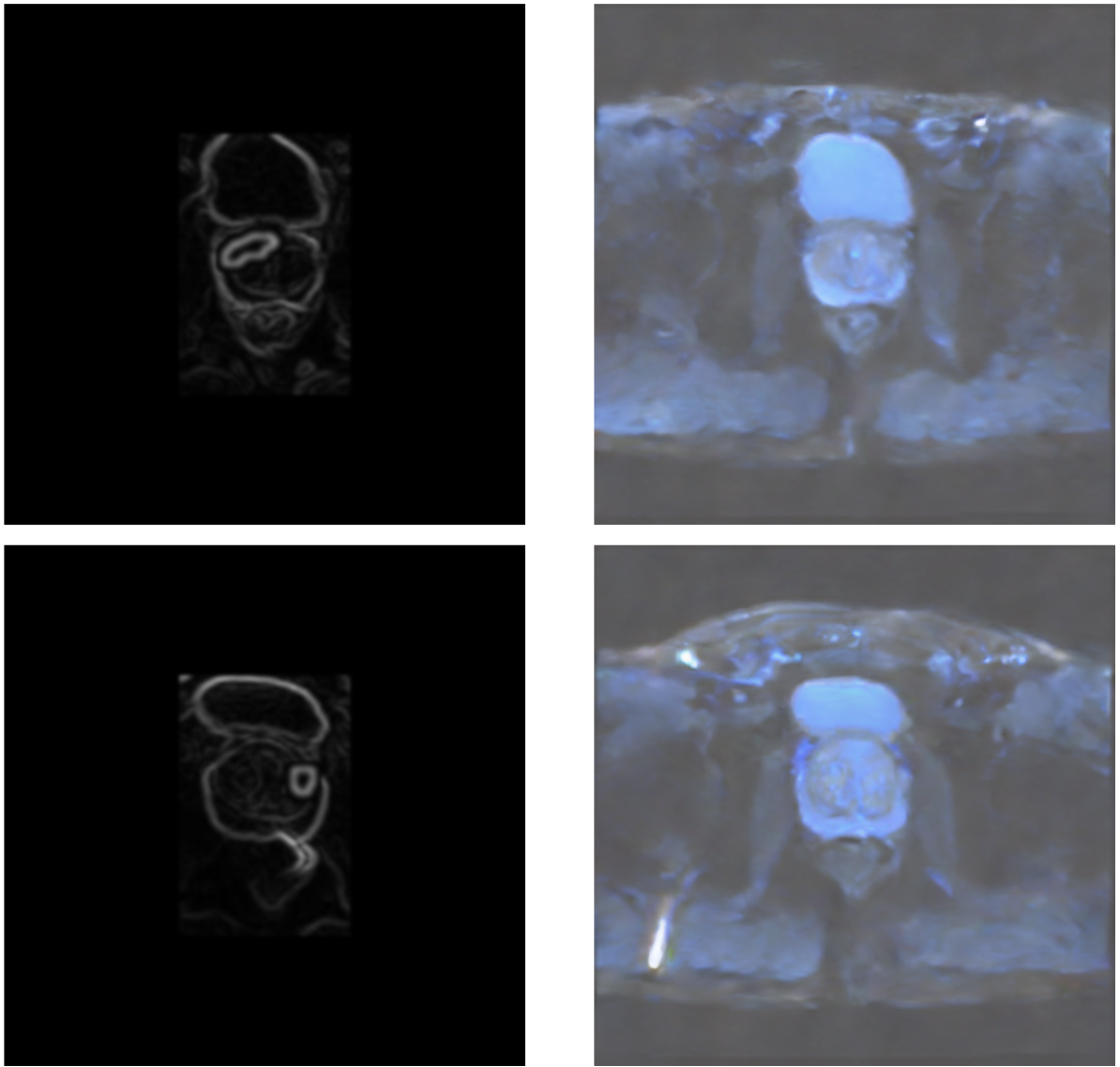}}
    \hfil
    \subfloat[Hand-drawn mask and DWI]{\includegraphics[width=.21\linewidth]{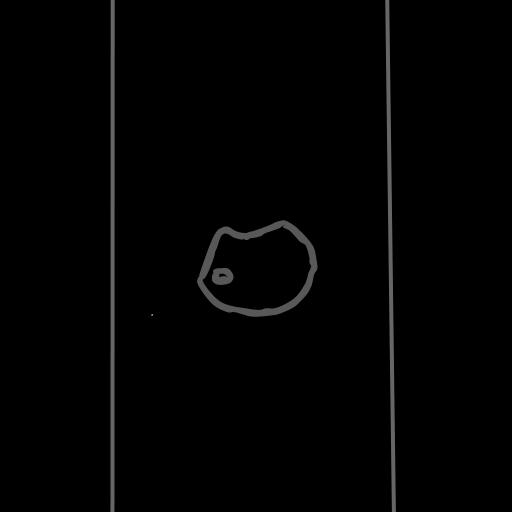} \includegraphics[width=.21\linewidth]{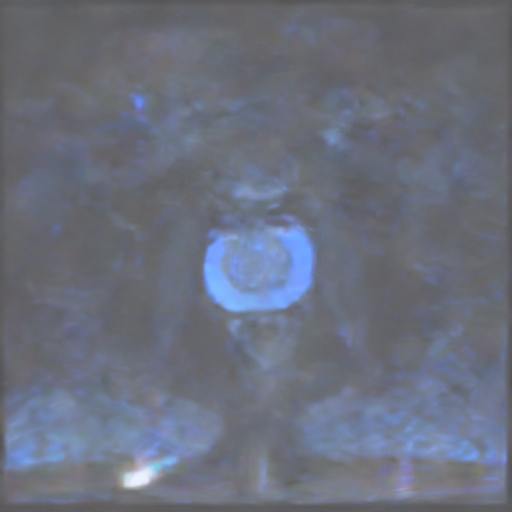}}
    \caption{Example DWI data produced by Anatomic-Conditional generation (a, b) and an example of a generated DWI from a hand-drawn conditioning mask (c).}
    \label{fig:example-generated-cases}
\end{figure}

\begin{figure}[!ht]
    \centering
    \includegraphics[width=\linewidth]{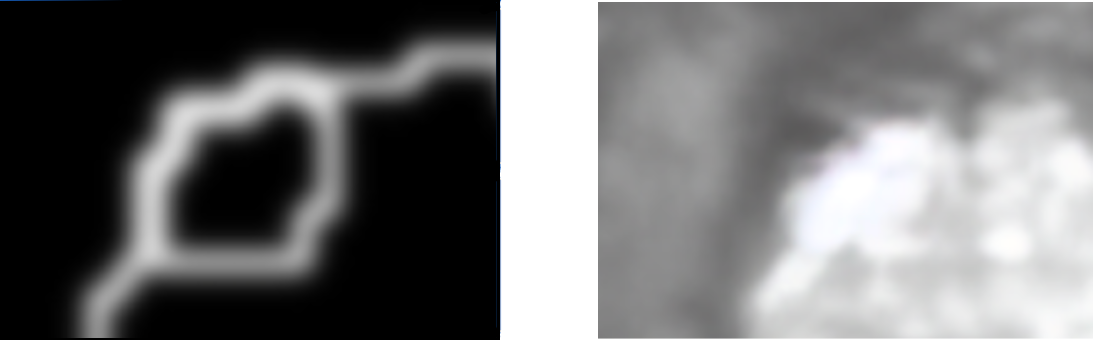}
    \caption{Example of an adjustment with the tumour region featuring a hyperintensity.}
    \label{fig:hyperintensity}
\end{figure}

We also use several conditioning strategies on our trained anatomic guidance networks to evaluate their generalization and applicability in the medical field. In particular, we investigate control schemes which offer flexibility and ease of use while still producing high quality scans. For the network trained on the single DWI channel (b=50), we used a simpler conditioning image that only features the prostate and lesion outlines (example seen in Figure~\ref{fig:example-generated-cases} (a)) and show that the network can still detect the anatomical components and control the output image accordingly. Specifically, the textual characteristics of the output image are adjusted with the tumour region featuring a hyperintensity similar to a real scan (shown in Figure~\ref{fig:hyperintensity}). However this strategy fails when implemented for the model that generates DWI images across all three channels (b=50, 400, 800). Subsequently, we use an alternate, more detailed conditioning scheme (shown in Figure~\ref{fig:process-view} where the control images are masked to the prostate and a soft edge map is generated using the conditioning image). We found that this technique allowed the model flexibility in inpainting the non-critical regions and better represented the varied real patient distribution as seen in Figure~\ref{fig:example-generated-cases} (b). However, this alternative strategy for three DWI channels requires the extra step of masks needing to be generated from a ground truth image, whereas the basic strategy just uses the outline of the prostate and lesion, which can easily be created manually using a drawing or annotation software, such as Microsoft Paint. An example of a generated image from a hand-drawn mask for the basic strategy can be seen in Figure~\ref{fig:example-generated-cases} (c).

\begin{figure}[!ht]
    \centering
    \includegraphics[width=\linewidth]{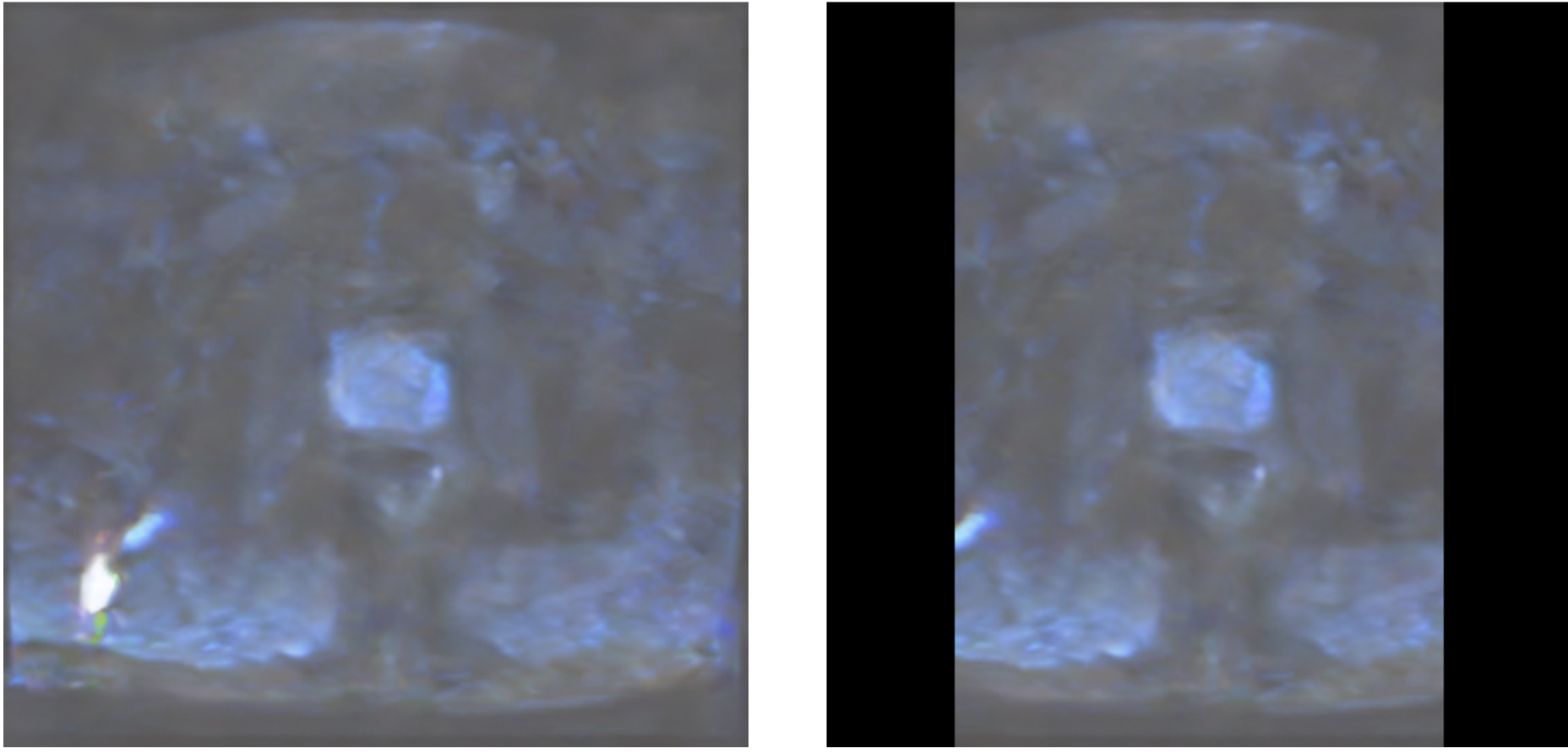}
    \caption{Example of a raw generated data sample (left) and the post-processed sample with side regions replaced with zeros (right).}
    \label{fig:comparison}
\end{figure}

We also evaluate the Fréchet Inception Distance (FID) \cite{heusel2018gans}, Learned Perceptual Image Patch Similarity (LPIPS) \cite{zhang2018unreasonable}, and Structural Similarity Index (SSIM) \cite{zhou2003ssim} between the training dataset and generated images for models trained on DWI (b=50) and DWI (b=50, 400, 800) and share the results in Table-\ref{table:metrics}. When performing evaluation, we generate 200 new images, each controlled using a unique conditioning image, and post-process by replacing the sides of the images with zeros to avoid having non prostate regions influence evaluation similarity scores. An example of a raw generated data sample and its post-processed sample is shown in Figure~\ref{fig:comparison}.

\begin{center}
  \begin{tabular}{cccc} \toprule
    {$Data$} & {$FID$} & {$LPIPS$} & {$SSIM$} \\ \midrule
    {$DWI (b=50)$}  & 85.52 & 0.2026 & 0.0029 \\
    {$DWI (b=50, 400, 800)$}  & 99.14  & 0.2180 & 0.0104 \\
    \bottomrule
  \end{tabular}
  \label{table:metrics}
\end{center}

\section{Conclusion}
In this paper, we introduce Cancer-Net PCa-Gen, an anatomic-conditional controlled latent diffusion strategy that generates realistic prostate DWI data. Leveraging the Cancer-Net PCa data, we provide experimental results with inputs of a single DWI channel and with inputs of three DWI channels and show that Cancer-Net PCa-Gen improves synthesis of prostate images with controlled tumour locations and better anatomical and textural fidelity. With these promising results, next steps include augmenting real patient data to improve neural network performance to supplement clinical support for prostate cancer. 

\begin{ack}
The authors thank the Natural Sciences and Engineering Research Council of Canada and the Canada Research Chairs Program.
\end{ack}

\bibliographystyle{unsrt}
\bibliography{neurips_2023}

\end{document}